\documentclass{iopart}
\usepackage{iopams}
\begin{document}

\title{Generalizations of some integrals over the unitary group}

\author{B. Schlittgen$^1$ and T. Wettig$^{1,2}$} 

\address{$^1$ Department of Physics, Yale University, New Haven, CT
  06520-8120, USA} 
\address{$^2$ RIKEN-BNL Research Center, Upton, NY, 11973-5000, USA}

\begin{abstract}
  Using the character expansion method, we generalize several
  well-known integrals over the unitary group to the case where
  general complex matrices appear in the integrand.  These integrals
  are of interest in the theory of random matrices and may also find
  applications in lattice gauge theory.
\end{abstract}




\section{Introduction}

Physical systems described by nonhermitian operators have recently
attracted a lot of attention in the literature.  Applications range
from flux-line pinning in superconductors \cite{MW96} over dissipation
and scattering in quantum chaos \cite{GHS88} to quantum chromodynamics
(QCD) at nonzero density \cite{Step96}.  In turn, the interest in
these applications has stimulated new mathematical studies of
nonhermitian matrices, in particular in the field of random matrix
theory.

We have been led to consider nonhermitian matrices in our work on the
color-flavor transformation \cite{Zirn96,Schl02,Budc02}.  This
transformation involves an integration over complex matrices $Z$
without any symmetry requirements.  Applying the transformation to
lattice QCD \cite{Budc02,Schl02b,Nonn02} results in a complex action
that is not amenable to standard Monte-Carlo algorithms. A
possible way of trying to resolve this problem is to write $Z=HU$ with $H$
hermitian and $U$ unitary, and to perform the integral over $U$
analytically so that only the integration over $H$ has to be done
numerically.  This approach leads to the following integral over the
unitary group,
\begin{equation}
\label{eq:I1}
  \mathcal{I}_1=\int\limits_{{\rm U}(N)}d\mu(U)\;{\det}^\nu U\;
  e^{\frac12\tr\,(AU+BU^\dagger)}\:,
\end{equation}
where $d\mu(U)$ is the Haar measure of U($N$), $\nu$ is an integer,
which without loss of generality we take to be non-negative, and
$A,B\in{\rm Gl}(N,\mathbb{C})$. The above integral is the main focus
of this paper.  It is well-known for $A=B^\dagger$
\cite{Brow81,Jack96,Akuz98} (where it leads to the effective partition
function of QCD in the Leutwyler-Smilga regime \cite{Leut92,Jack96}),
but to the best of our knowledge $\mathcal{I}_1$ had not been computed
before for $A\ne B^\dagger$.  We found that the latter case can be
solved rather easily using the character expansion method put forward
by Balantekin \cite{Bala84,Bala00}.  Although the case of non-integer
$\nu$ appears to be beyond this approach, we expect our result to hold
in that case as well. Employing the same method, we could also compute
the integral which arises from the factorization $Z = U\Lambda V^{\dag}$
in the color-flavor transformed action mentioned above. The matrices
$U$ and $V$ are unitary, while $\Lambda$ is diagonal with nonnegative
entries. The 
resulting integral in slightly generalized form is given by
\begin{equation}
\label{eq:I2}
  \mathcal{I}_2=\int\limits_{{\rm U}(N)}d\mu(U)\;
      \int\limits_{{\rm U}(N)}d\mu(V)\;
      {\det}^\nu (UV)\;
      e^{\frac12\tr\,(UAVB+CV^\dagger DU^\dagger)}\:,
\end{equation}
where $A,B,C,D\in{\rm Gl}(N,\mathbb{C})$.  This integral was
previously known only for the case of $\nu=0$, $C=B^\dagger$ and
$D=A^\dagger$ \cite{Bere58,Guhr96,Jack96}.
We also show that the generalization of $\mathcal{I}_2$ to the case
of unequal dimensions of $U$ and $V$ leads to an integral which can
be non-zero only if determinants of $U$ and $V$ are absent from the integrand.
We conjecture an expression for the result in this case.

In Sec.~\ref{sec:calculation}, we derive results for $\mathcal{I}_1$
and $\mathcal{I}_2$. In addition, we discuss $\mathcal{I}_2$ with
unequal dimensions of $U$ and $V$ and briefly consider the (trivial)
generalization of the Itzykson-Zuber integral to the case of general
complex matrices in the integrand.  We also comment on the
applicability of our results to the case of noninvertible matrices.  
Conclusions are drawn in Sec.~\ref{sec:conlusions}.

\section{Calculation of the group integrals}
\label{sec:calculation}

Our calculations rely on the formalism of Ref.~\cite{Bala00}
and on several examples worked out in detail therein.  Rather than
reproducing the material from that work, we will refer to the
relevant equations in Ref.~\cite{Bala00} whenever appropriate.  Thus, the
reader is advised to have a copy of this reference at hand.

The key observation for the generalization of the results of Ref.~\cite{Bala00}
is that the representation theories of the groups U($N$) and Gl($N$) are
essentially the same. In particular, Weyl's character formula
\begin{equation}
\chi_r (X) = \frac{\det \left(x_i^{n_j +N-j}\right)}{\Delta\left( x_1,
\ldots,x_N\right)},
\end{equation}
holds not just for unitary, but also for general linear matrices,
cf.\ Ref.~\cite{Hua63}. Here, $r=(n_1,\ldots,n_N)$ denotes an
irreducible representation
of Gl($N$) labelled by non-negative, non-increasing integers. The $x_i$
($i=1,\ldots,N$) are the eigenvalues of the matrix $X$.

\subsection{Calculation of $\mathcal{I}_1$}

Using Eq.~(3.5) of Ref.~\cite{Bala00}, we write 
\begin{eqnarray}
  {\det}^\nu (AU)\;e^{\tr\,AU}=\sum_r \alpha_r^{(\nu)}\chi_r(AU)\;,\\
  e^{\tr\,BU^\dagger}=\sum_r \alpha_r^{(0)}\chi_r(BU^\dagger)\:.
\end{eqnarray}
Here, the sums are over all irreducible representations of Gl($N$)
labeled by $r = (n_1,n_2,\ldots,n_N)$ as above.  The corresponding
characters are denoted by $\chi_r$.  For convenience, we have left out
the factor of $\frac12$ in the exponent of Eq.~(\ref{eq:I1}), which
will be reinstated in the final result.  The coefficients in the
character expansion are given by
\begin{equation}
\label{eq:alphanu}
  \alpha_r^{(\nu)}=\det\left[\frac1{(n_j-\nu+i-j)!}\right]\:,
\end{equation}
where $i$ and $j$ run from 1 to $N$, labelling the rows and columns of the 
matrix. We thus obtain
\begin{eqnarray}
\label{eq:I1hat}
  \hat\mathcal{I}_1&\equiv\int\limits_{{\rm U}(N)}d\mu(U)\;
  {\det}^\nu (AU) \; e^{\tr\,(AU+BU^\dagger)}\nonumber\\
  &=\sum_r\sum_{r'}\alpha_r^{(\nu)}\alpha_{r'}^{(0)}
  \int\limits_{{\rm U}(N)}d\mu(U)\;
  \chi_r(AU)\,\chi_{r'}(BU^\dagger).
\end{eqnarray}
Now, we can write
\begin{equation}
\label{eq:charsep}
\chi_r(AU)\chi_{r'}(BU^{\dagger}) = A^{(r)}_{ab}U^{(r)}_{ba}
B^{(r')}_{cd}U^{(r')*}_{cd},
\end{equation}
where the superscript serves as a reminder that these matrices
live in the representations $r$ and $r'$ of Gl($N$), respectively.
If we restrict ourselves to unitary matrices, we
have corresponding irreducible representations of the subgroup
U($N$) of Gl($N$), and hence, we can use the group theoretical result
\begin{equation}
\label{eq:ortho}
  \int\limits_{{\rm U}(N)}d\mu(U)\;U^{(r)}_{ba}\,U^{(r')*}_{cd}
  =\frac1{d_r}\delta^{rr'}\delta_{bc}\delta_{ad}\:,
\end{equation}
where $d_r$ is the dimension of $r$, regardless of whether we consider it
as a representation of Gl($N$) or U($N$). It then follows that
\begin{equation}
\label{eq:I1hat2}
  \hat{\mathcal{I}}_1
  =\sum_r \frac{\alpha_r^{(0)}}{d_r}\,\alpha_{r}^{(\nu)}\, \chi_r(AB)\:.
\end{equation}
From Eqs.~(3.5), (3.3) and (2.9) of Ref.~\cite{Bala00}, we obtain
\begin{equation}
\label{eq:alphanud}
  \frac{\alpha_{r}^{(\nu)}}{d_r}=\prod_{i=1}^N\frac{(N-i)!}{(k_i-\nu)!}
  \qquad{\rm with}\;\; k_i=N+n_i-i\:.
\end{equation}
The matrix $AB$ has $N$ eigenvalues which we denote by
$\mu_1^2,\ldots,\mu_N^2$.  We now apply Weyl's formula, 
\begin{equation}
\label{eq:weyl}
  \chi_r(AB)=\frac{\det\left[\mu_i^{2(n_j+N-j)}\right]}
  {\Delta(\mu^2)}\:,
\end{equation}
where
\begin{equation}
  \Delta(x)=\prod_{i<j}^N(x_i-x_j)
\end{equation}
is the Vandermonde determinant.  Inserting this expression, together
with Eqs.~(\ref{eq:alphanu}) and (\ref{eq:alphanud}), into
Eq.~(\ref{eq:I1hat2}) yields
\begin{equation}
  \hat\mathcal{I}_1=\left[\prod_{n=1}^{N-1} n!\right]\:\frac1{\Delta(\mu^2)}\;
  \sum_r \det \left[\frac1{k_j!(k_j-N-\nu+i)!}\right]
  \det\Bigl[\mu_i^{2k_j}\Bigr]\:.
\end{equation}
Applying the Binet-Cauchy formula, see Eq.~(B4) of Ref.~\cite{Bala00},
and using the power series expansion of the Bessel function,
\begin{equation}
\label{eq:bessel}
  \frac{I_\lambda(2y)}{y^\lambda}=\sum_{k=0}^\infty
  \frac{y^{2k}}{k!(k+\lambda)!} \:,
\end{equation}
we obtain immediately
\begin{equation}
  \hat\mathcal{I}_1=\left[\prod_{n=1}^{N-1} n!\right]\:\frac1{\Delta(\mu^2)}
  \;\det\Bigl[\mu_j^{N-i+\nu}I_{i-N-\nu}(2\mu_j)\Bigr]\:.
\end{equation}
Pulling out the factors of $\mu_j^\nu$ and rearranging the determinant
using $I_n=I_{-n}$ yields
\begin{eqnarray}
  \det\Bigl[\mu_j^{N-i+\nu}I_{i-N-\nu}(2\mu_j)\Bigr]
  &={\det}^{\frac\nu2}(AB)\,
  \det\Bigl[\mu_j^{N-i}I_{i-N-\nu}(2\mu_j)\Bigr]\nonumber\\
  &={\det}^{\frac\nu2}(AB)\,
  \det\Bigl[\mu_j^{i-1}I_{\nu+i-1}(2\mu_j)\Bigr]\:.
\end{eqnarray}
Finally, we rescale $A$ and $B$ by $\frac12$ to reinstate the factor
of $\frac12$ in the exponent of Eq.~(\ref{eq:I1}) to obtain
\begin{eqnarray}
  \fl\int\limits_{{\rm U}(N)}d\mu(U)\;{\det}^\nu U \;
  e^{\frac12\tr\,(AU+BU^\dagger)}\nonumber\\
  \lo=2^{\frac{N(N-1)}2}\left[\prod\limits_{n=1}^{N-1}n!\right]
  \left(\frac{\det B}{\det A}\right)^{\!\!\frac\nu2}\;
  \frac{\det\left[\mu_i^{j-1} I_{\nu+j-1}(\mu_i)\right]}
  {\Delta(\mu^2)}\:,
  \label{eq:i1}
\end{eqnarray}
where, once again, the $\mu_i^2$ are the eigenvalues of $AB$.

\subsection{Calculation of $\mathcal{I}_2$}

Note first that in the integrand of $\mathcal{I}_2$, the determinants of 
$U$ and $V$ have to be raised to the same power, as indicated in 
Eq.~(\ref{eq:I2}); otherwise the integrations over the U(1) subgroups 
of $U$ and $V$ simply render $\mathcal{I}_2$ zero.

Using again Eq.~(3.5) of Ref.~\cite{Bala00}, we have
\begin{eqnarray}
  {\det}^\nu(UAVB)\;e^{\tr\,UAVB}=\sum_r
  \alpha_r^{(\nu)}\chi_r(AUVB)\;,\\ 
  e^{\tr\,CV^\dagger DU^\dagger}=\sum_r
  \alpha_r^{(0)}\chi_r(CV^\dagger DU^\dagger)
\end{eqnarray}
with $\alpha_r^{(\nu)}$ given in Eq.~(\ref{eq:alphanu}).  Thus,
\begin{eqnarray}
\label{eq:I2hat}
\fl \begin{array}{rl}
{\displaystyle \hat\mathcal{I}_2}
&{\displaystyle \equiv\int\limits_{{\rm U}(N)}d\mu(U)\;
  \int\limits_{{\rm U}(N)}d\mu(V)\;
  {\det}^\nu (UAVB)\; e^{\tr\,(UAVB+CV^\dagger DU^\dagger)}} \\
&{\displaystyle =\sum_r\sum_{r'}\alpha_r^{(\nu)}\alpha_{r'}^{(0)}
  \int\limits_{{\rm U}(N)}d\mu(U)\;\int\limits_{{\rm U}(N)}d\mu(V)\;
  \chi_r(UAVB)\,\chi_{r'}(CV^\dagger DU^\dagger)}\\
&{\displaystyle =\sum_r \frac{\alpha_r^{(\nu)}\alpha_{r}^{(0)}}{d_r}
  \int\limits_{{\rm U}(N)}d\mu(U)\; \chi_r(BUADU^\dagger C)}\\
&{\displaystyle =\sum_r \frac{\alpha_r^{(\nu)}\alpha_{r}^{(0)}}{d_r^2}
  \chi_r(AD)\chi_r(BC)\:,}
\end{array}
\end{eqnarray}
where we have made use of Eqs.~(\ref{eq:charsep}) and (\ref{eq:ortho}).
From Eq.~(\ref{eq:alphanud}) we have
\begin{equation}
  \frac{\alpha_r^{(\nu)}\alpha_{r}^{(0)}}{d_r^2}=\left[
  \prod_{n=1}^{N-1}n!\right]^2\;\prod_{i=1}^N\frac1{k_i!(k_i-\nu)!}\:.
\end{equation}
Now denote the eigenvalues of the matrices $AD$ and $BC$ by
$x_1^2,\ldots,x_N^2$ and $y_1^2,\ldots,y_N^2$, respectively.  Using
again Weyl's formula (\ref{eq:weyl}), Eq.~(\ref{eq:I2hat}) thus
becomes
\begin{equation}
  \hspace*{-15mm} 
  \hat\mathcal{I}_2=\left[\prod_{n=1}^{N-1}n!\right]^2
  \frac1{\Delta(x^2)\Delta(y^2)} \sum_r 
  \prod_{i=1}^N\frac1{k_i!(k_i-\nu)!}
  \det\Bigl[x_i^{2k_j}\Bigr]\det\Bigl[y_i^{2k_j}\Bigr]\:.
\end{equation}
Employing the expansion theorem given in Eq.~(B2) of
Ref.~\cite{Bala00} and noting Eq.~(\ref{eq:bessel}) again, this yields
\begin{equation}
  \hat\mathcal{I}_2=\left[\prod_{n=1}^{N-1}n!\right]^2
  \frac1{\Delta(x^2)\Delta(y^2)}
  \det\left[f(x_iy_j)\right]
\end{equation}
with
\begin{equation}
  f(z)=z^\nu\,I_{-\nu}(2z)\:.
\end{equation}
Pulling the factors of $x_i^\nu$ and $y_j^\nu$ out of the determinant
and using $I_n=I_{-n}$, we obtain
\begin{equation}
  \hat\mathcal{I}_2=\left[\prod_{n=1}^{N-1}n!\right]^2
  {\det}^{\frac\nu2}(ABCD)\,
  \frac{\det\left[I_\nu(2x_iy_j)\right]}{\Delta(x^2)\Delta(y^2)}\:.
\end{equation}
We finally rescale $A$, $B$, $C$ and $D$ by $1/\sqrt2$ to reinstate
the factor of $\frac12$ in the exponent of Eq.~(\ref{eq:I2}) to obtain
\begin{eqnarray}
  \fl\int\limits_{{\rm U}(N)}d\mu(U)\;
  \int\limits_{{\rm U}(N)}d\mu(V)\; {\det}^\nu (UV)\;
  e^{\frac12\tr\,(UAVB+CV^\dagger DU^\dagger)}\nonumber\\
  =2^{N(N-1)}\left[\prod_{n=1}^{N-1}n!\right]^2
  \left(\frac{\det(CD)}{\det(AB)}\right)^{\!\!\frac\nu2}
  \frac{\det\left[I_\nu(x_iy_j)\right]}{\Delta(x^2)\Delta(y^2)}\:.
\label{I2res}
\end{eqnarray}
Once again, the $x_i^2$ and $y_j^2$ are the eigenvalues of $AD$ and
$BC$, respectively.

Let us now consider the case in which $U$ and $V$ have different
dimensions, i.e.,
\begin{equation}
\mathcal{I}_2^{(N,M)}=\int\limits_{{\rm U}(N)}d\mu(U)\;
  \int\limits_{{\rm U}(M)}d\mu(V)\; 
  e^{\frac{1}{2}\tr\,(UAVB+CV^\dagger DU^\dagger)}.
\end{equation}
In this case, $A$ and $C$ are complex $N\times M$ matrices,
and $B$ and $D$ are complex $M\times N$ matrices. For definiteness,
we shall take $M<N$. 

At this point, we have not been able to prove a result for
$\mathcal{I}_2^{(N,M)}$, but we conjecture,
based on explicit calculations for small $N$ and $M$, as well as 
on numerical experimentation, that the result takes the form
\begin{equation} 
\fl \mathcal{I}_2^{(N,M)}= 2^{M(N-1)} \left[\prod_{n=N-M}^{N-1} n! \right]
\left[\prod_{m=N-M}^{M-1}m!\right] 
\frac{\det\left[I_{N-M}(x_i y_j)\right]}{\Delta(x^2)
\Delta(y^2)
\prod_{i=1}^{M}(x_i y_i)^{N-M}}\: .
\end{equation}
Here, $x_i^2$ and $y_i^2$ ($i=1,\ldots,M$) denote the (non-zero) eigenvalues of
$DA$ and $BC$, respectively. This expression also reduces to the well-known
result in the case where $C = B^{\dag}$ and $D = A^{\dag}$, cf.\
Refs.~\cite{Bere58,Jack96,Guhr96}.

Note that we have not included any determinant terms in the integrand of 
$\mathcal{I}_2^{(N,M)}$. If we included, say, 
$\det^{\nu} U\, \det^{\eta} V$ in the integrand, integrations 
over the U($1$) subgroups of U($N$) and U($M$) show immediately
that the value of the resulting integral is zero unless, possibly,
$\nu$ and $\eta$ are related by $N\nu=M\eta$. We now show
that, even if this relation holds, the integral gives zero for any
$\nu\neq 0$, and hence also for $\eta\neq 0$. To see this, suppose
that $\nu\neq 0$, and let us perform the integral of $U$ over U($N$),
leaving the integral over U($M$) untouched for the moment. The result
could be read off from Eq.~(\ref{eq:i1}) if the matrices
$AVB$ and $CV^{\dag}D$ were Gl($N$) matrices. However, since $M<N$,
these matrices are not of full rank, and therefore $N-M$ of their
eigenvalues are equal to zero. Except on a set of measure zero they have the 
same rank, so that a limiting process leads to a finite value of 
$\det(CV^{\dag}D)/\det(AVB)$, given by $\det(V^{\dag}DC)/\det(VBA)$. 
We then need to find the limit of 
$\left.\det\left[\mu_i^{j-1}I_{\nu+j-1}(\mu_i)\right]\right/\Delta(\mu^2)$
as $\mu_{M+1},\ldots,\mu_N \to 0$. In this context, $\mu_1^2,\ldots, \mu_M^2$
denote the non-zero eigenvalues of $AVBCV^{\dag}D$.
In fact, letting just $\mu_N\to 0$,
it is easy to see that the above expression goes to zero, unless
$\nu =0$, which shows that $\mathcal{I}_2^{(N,M)}=0$, unless
$\nu=\eta=0$.

\subsection{Generalization of the Itzykson-Zuber integral}

The integral
\begin{equation}
  \mathcal{I}_3=\int\limits_{{\rm U}(N)}d\mu(U)\;e^{\tr\,(AUBU^\dagger)}
  =\left[\prod_{n=1}^{N-1}n!\right]\:\frac{\det\left[\exp(x_iy_j)\right]}
  {\Delta(x)\Delta(y)}
\label{I3res}
\end{equation}
was computed in Ref.~\cite{Itzy80} for the case where $A$ and $B$ are
hermitian matrices with real eigenvalues $x_1,\ldots,x_N$ and
$y_1,\ldots,y_N$, respectively.  This is a special case of a more
general result due to Harish-Chandra \cite{Hari58}.

Following the calculation of this integral in Ref.~\cite{Bala00},
it is immediately obvious that the only change in the final result is
the replacement of the eigenvalues of the hermitian matrices $A$ and
$B$ by the eigenvalues of their general complex versions.

Also, including the determinant of $U$ in the integrand (raised to any
nonzero power) would give zero due to the integration over the U(1)
subgroup.

\subsection{Comment on noninvertible matrices}

In deriving the above results, we have
assumed that the matrices $A$, $B$, $C$ and $D$ are elements of
Gl($N,\mathbb{C}$).  However, the integrals $\mathcal{I}_1$ through
$\mathcal{I}_3$ exist even if the matrices on which they depend are
not of full rank.  In this case, we can consider the limit in which
one or more of the eigenvalues of the matrix approach zero.  Uniform
convergence permits the interchange of this limit and the integration
over the unitary group.  A l'H\^opital procedure on the right-hand
side of Eq.~(\ref{eq:i1}), (\ref{I2res}) or (\ref{I3res}) then leads
to a finite (though possibly zero) result.

\section{Conclusions}
\label{sec:conlusions}

We have derived generalizations of several well-known integrals over
the unitary group to the case where general complex matrices appear in
the integrand.  These integrals may find applications in lattice gauge
theory but are also of purely mathematical interest, in particular in
the theory of random matrices.  As mentioned in the introduction, our
motivation for studying these integrals originated from the complex
action problem that arises if the color-flavor transformation is
applied to lattice QCD.  We found that the integral $\mathcal{I}_1$
solved this complex action problem for one quark flavor but
unfortunately not for two or more flavors.

Our results were obtained by a straightforward application of
Balantekin's character expansion method.  It would be interesting to
investigate the feasibility of other well-known methods to compute
integrals over the unitary group, such as the diffusion equation
method, in the cases we have considered.  It should also be possible
to generalize the results of the present paper to integrals over the
super-unitary group.

\section*{Acknowledgments}

This work was supported in part by the U.S.\ Department of Energy
under contract No.\ DE-FG02-91ER40608 and in part under the ECT*
'STATE' contract.  We thank T. Guhr for interesting conversations.

\end{document}